\documentclass[namedreferences]{solarphysics}
\usepackage[optionalrh,solaenum]{spr-sola-addons} 
\usepackage{graphicx}                    
\usepackage{color}                       
\usepackage{url}                         
\usepackage{lscape}						 

\usepackage{multirow}					 

\newcommand{\etal}{{\it et al.}}
\newcommand{\insitu}{{\it in-situ}}

\begin{document}

\begin{article}

\begin{opening}

\title{Three-dimensional evolution of erupted flux ropes from the Sun (2--20~$R_{\odot}$) to 1~AU}

%
\author{A.~\surname{Isavnin}$^{1}$\sep
        A.~\surname{Vourlidas}$^{2}$\sep
        E.K.J.~\surname{Kilpua}$^{1}$
       }

%
\runningauthor{A. Isavnin, {\etal}}
\runningtitle{Three-dimensional evolution of erupted flux ropes from the Sun (2--20~$R_{\odot}$) to 1~AU}

%
  \institute{$^{1}$ Department of Physics, University of Helsinki, P.O.Box 64, FI-00014, Finland, email: \url{Alexey.Isavnin@helsinki.fi}\\
  			 $^{2}$ Space Science Division, Naval Research Laboratory, Washington DC, USA
             }

\begin{abstract}
Studying the evolution of magnetic clouds entrained in coronal mass ejections using {\insitu} data is a difficult task since only a limited number of observational points is available at large heliocentric distances. Remote sensing observations can, however, provide important information for events close to the Sun. In this work we estimate the flux rope orientation first in the close vicinity of the Sun (2--20~$R_{\odot}$) using forward modeling of STEREO/SECCHI and SOHO/LASCO coronagraph images of coronal mass ejections and then {\insitu} using Grad-Shafranov reconstruction of the magnetic cloud. Thus, we are able to measure changes in the orientation of the erupted flux ropes as they propagate from the Sun to 1~AU. We present both techniques and use them to study 15~magnetic clouds observed during the minimum following Solar Cycle~23 and the rise of Solar Cycle~24. This is the first multievent study to compare the three-dimensional parameters of CMEs from imaging and {\insitu} reconstructions. The results of our analysis confirm earlier studies showing that the flux ropes tend to deflect towards the solar equatorial plane. We also find evidence of rotation on their travel from the Sun to 1~AU. In contrast to past studies, our method allows one to deduce the evolution of the three-dimensional orientation of individual flux ropes rather than on a statistical basis.
\end{abstract}

%

\end{opening}

%

\section{Introduction}\label{s:intro} 

Coronal mass ejections (CMEs) are massive bursts of plasma and magnetic field from the Sun into the interplanetary space. Interplanetary coronal mass ejections (ICMEs) are the heliospheric counterparts of the CMEs. ICMEs are one the main drivers of space weather (\opencite{Tsurutani1988}, Huttunen {\etal}, \citeyear{Huttunen2002}, \opencite{Zhang2007}). ICMEs show a variety of signatures in {\insitu} observations at 1~AU. Some of them such as the enhancement of the magnetic field, smooth monotonic rotation of the magnetic field through a large angle, low proton temperature and low plasma $\beta$, indicate the existence of a flux rope structure within the body of the ICME. ICMEs with embedded magnetic flux ropes are called magnetic clouds (MCs). The list of signatures found in {\insitu} measurements of MCs can be found in the paper of \inlinecite{Zurbuchen2006}. Approximately one-third of all ICMEs observed at 1~AU show magnetic cloud signatures \cite{Gosling1990}, but there are indirect evidences that all ICMEs might have central flux ropes. \inlinecite{Krall2007} conluded after analyzing \textit{Solar Maximum Mission} (SMM) coronagraph data  that all CMEs in his sample may share a similar hollow-flux-rope structure (see also Vourlidas {\etal}, 2013, this volume). As suggested by \inlinecite{Jian2006}, the absence of MC signatures in many ICMEs may be a positional effect: their analysis showed that in about two-thirds of the cases the spacecraft encounters the ICME so far from the center that the central flux rope is not identifiable. Since the southward magnetic field has the strongest influence on the Earth's magnetosphere, the geoeffectivity of MCs depends on their orientation. The orientation of the MC in the low solar corona and at 1~AU can differ greatly, so it is crucial to understand how the orientation of MCs changes on their journey from the Sun to 1~AU. This knowledge will help us to better understand the processes affecting CME propagation in the heliosphere, improve the solar and heliosphere magnetohydrodynamics simulation techniques and hence improve space weather forecasting.

The orientation of MCs during their journey from the Sun to 1~AU can change significantly. The change of the orientation can be decomposed into latitudinal deflection, longitudinal deflection and rotation. The latitudinal deflection of CMEs in the low corona was reported already by MacQueen {\etal} (\citeyear{MacQueen1986}). According to their work, CMEs tend to deflect towards the solar equatorial plane by about 2 degrees on average while travelling from $2R_{\odot}$ to $4R_{\odot}$ during solar minimum. \inlinecite{Plunkett2001} found that the latitude distribution of CMEs in the outer corona near solar minimum is very different to the distribution in the inner corona, suggesting that the propagation of CMEs in the inner corona is controlled by the large-scale solar magnetic field which tends to push the CME towards the equatorial streamer belt as it propagates outward. On the other hand, CMEs show no average latitudinal deflection during solar maximum. Cremades {\etal} (\citeyear{Cremades2006}) has shown that the deviation of CMEs with respect to their source regions is always equatorward near solar minimum, while deviations to higher latitudes are also frequent during solar maximum. They also found a significant correlation of the deviation with the number of coronal holes, their area and their distance to the CME source regions. \inlinecite{Kilpua2009} pointed that the spatial dimensions of the erupting CME might play a role in determining whether a CME will be deflected towards the equator, \textit{i.e.} slower CMEs with wide longitudinal extent could not penetrate through the background coronal fields.

The evidence for longitudinal deflection is based on the statistics of the solar source regions of geoeffective halo CMEs (\textit{i.e.} CMEs that arrived to the Earth and produced geomagnetic storms). It was found by \inlinecite{Wang2002} that the East-West distribution of the CME sources is asymmetrical -- the number of geoeffective halo CMEs originating from the western hemisphere is larger by 57\% compared to the ones originating from the eastern hemisphere. In the paper by \inlinecite{Wang2011} the CME deflections were classified into three types: (i) asymmetrical expansion of the CME, (ii) nonradial ejection and (iii) deflected propagation. Deflected propagation in the lower corona can be caused by the interaction of the CME with other neighbouring large-scale magnetic field structures such as coronal holes.

MCs can also experience rotation on their journey to 1~AU. The possible reason for MC rotation can be the interaction with neighbouring magnetic field structures and the kink instability \cite{Torok2003}. But CME rotation is not expected at large distances when the ambient magnetic field is weak \cite{Lynch2009}. Sometimes MCs can suffer rather rapid rotation, however. \inlinecite{Vourlidas2011} reported a MC rotating at a rate of $60^{\circ}$ per day in the low corona. Based on statistical evidence MCs seem to rotate towards the heliospheric current sheet (HCS) so that the MC stays aligned with the local HCS as shown by \inlinecite{Yurchyshyn2008} and Yurchyshyn {\etal} (\citeyear{Yurchyshyn2009}).

It is difficult to study the evolution of MCs from the Sun to 1~AU, since there is only a limited number of observing points for analysis of such a complicated and large structure. The launch of the STEREO mission \cite{Kaiser2008} made it possible to obtain stereoscopic coronagraph images of the Sun which gave rise to forward modeling (Thernisien {\etal}, \citeyear{Thernisien2009}, \citeyear{Thernisien2011}). Paired  with {\insitu} measurements it allows to obtain the geometrical parameters of MCs in different temporal and spatial stages of their evolution. In this work we study the evolution of MCs registered during the minimum following Solar Cycle~23 and the rise of Solar Cycle~24. We use forward modeling technique to study the properties of MCs close to the Sun (5--20~$R_{\odot}$) and Grad-Shafranov reconstruction at 1~AU. 

Our paper is organized as follows: in Section 2 we will describe our method of estimating the orientation of MCs using one example from our list of 15 events, in Section 3 we present the results of our analysis and discuss them in Section 4.

\section{Methodology using an example event}\label{s:method}

A coronagraph is a telescope pointed at the Sun with an occulter blocking the solar disk light. The solar disk is about $10^{6}-10^{9}$ times brighter than the inner corona, thus the occulter is necessary to observe the solar corona with sufficient contrast to reveal faint structures. There are three coronagraphs on board the SOHO spacecraft with fields of view of 1.1--3~$R_{\odot}$ (C1), 2.2--6~$R_{\odot}$ (C2) and 3.5--30~$R_{\odot}$ (C3), part of the LASCO (\textit{Large Angle and Spectrometric Coronagraph}) experiment \cite{Brueckner1995}. The STEREO spacecraft are supplied with two coronagraphs each with fields of view of 1.5--4~$R_{\odot}$ (COR1) and 2.5--15~$R_{\odot}$ (COR2) and a heliospheric imager (HI) pointed towards the Sun--Earth line, part of the SECCHI (\textit{Sun Earth Connection Coronal and Heliospheric Investigation}: \inlinecite{Howard2008}) package. In our study we tracked each CME from 2 to 20--40~$R_{\odot}$ using LASCO C2 and C3 and SECCHI COR2 and HI telescopes. We required that each CME was captured {\insitu} by at least one spacecraft (STEREO and/or \textit{Wind}).

\begin{figure}
\centerline{\includegraphics[width=1.0\textwidth,clip=]{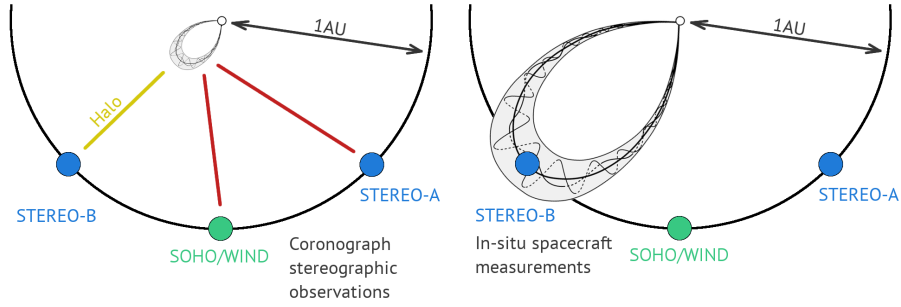}}
\caption{Scheme of the event suitable for our analysis.}\label{fig:event_scheme}
\end{figure}

The method we developed in this paper to study the evolution of a MC is based on the forward modeling of coronagraph data \cite{Thernisien2009} and Grad-Shafranov reconstruction of {\insitu} MCs \cite{Hu2002}. We present this method alongside an example event of the CME on 4 November 2010 and the associated {\insitu} MC measured on 8 November 2010 by the STEREO-B spacecraft. The schematic representation of this event is depicted on Figure~\ref{fig:event_scheme}.

\begin{figure}
\centerline{\includegraphics[width=1.0\textwidth,clip=]{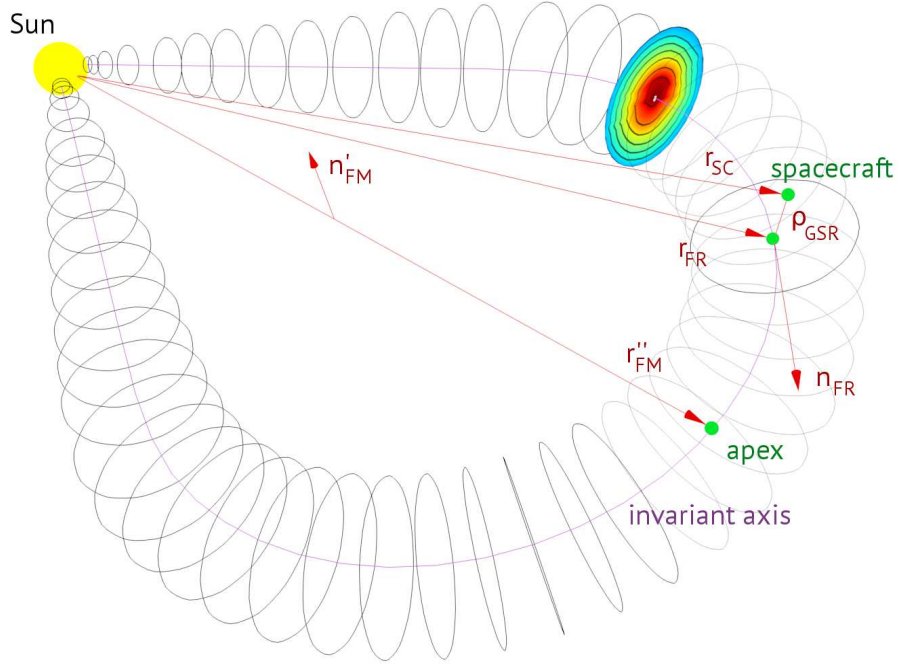}}
\caption{Three-dimensional scheme of magnetic cloud. $\mathbf{\hat{r}''}_{\mathrm{FM}}$ is the global axis of the flux rope, $\mathbf{\hat{n}}_{\mathrm{FR}}$ is the local invariant axis of the flux rope at the point of intersection with the spacecraft, $\rho_{\mathrm{GSR}}$ is the impact parameter.}\label{fig:mc_scheme}
\end{figure}

We assume that the invariant axis of a MC lies in a single plane (Figure~\ref{fig:mc_scheme}), so that when the MC changes its orientation, the plane containing the invariant axis changes its orientation accordingly. We will refer to it as the MC plane from now on. The orientation of the MC can be characterised by the normal to the MC plane and the direction from the Sun to the apex of the MC (Figure~\ref{fig:mc_scheme}).

\begin{figure}
\centerline{\includegraphics[width=1.0\textwidth,clip=]{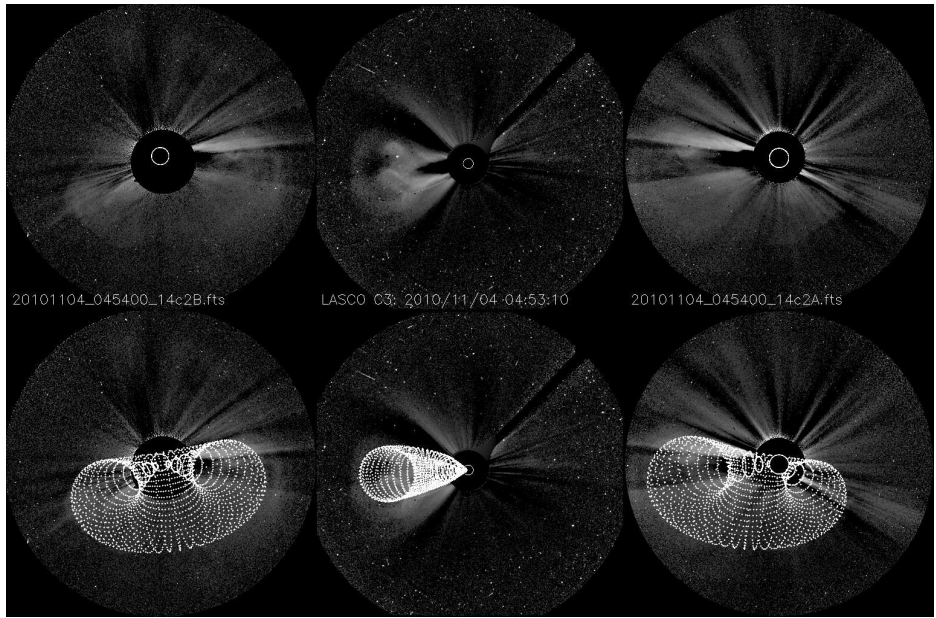}}
\caption{Example of forward modeling (lower panels) of the CME of 4 November 2010 (upper panels). The gray-scale images from left to right are the coronagraph images obtained by STEREO-B/SECCHI COR2, SOHO/LASCO C3 and STEREO-A/SECCHI COR2 telescopes.}\label{fig:fm_example}
\end{figure}

The multi-spacecraft forward modeling (FM) was introduced by \inlinecite{Thernisien2009} and is implemented in the SolarSoft package. The technique is based on the fitting of a three-dimensional hollow-croissant-shaped structure to stereoscopic coronagraph images of a CME. It is preferable to use all three spacecraft observations for a fitting procedure since at least one of them will observe the CME as halo or partial halo resulting in faint coronagraph images. Figure~\ref{fig:fm_example} shows the FM fitting for the CME on 4 November 2010. The separation between the STEREO spacecraft was about $166^{\circ}$, so both STEREO spacecraft observed partial halo CME (backside partial halo CME for STEREO-A). LASCO, at the same time, observed the CME on the Eastern part of the limb. The output of FM that we are interested in are the direction of radial CME propagation $(\theta_{\mathrm{FM}},\phi_{\mathrm{FM}})$ and its rotation angle $\gamma_{\mathrm{FM}}$. These outputs are given in the Stonyhurst coordinate system \cite{Thompson2006}. The direction of CME propagation $\mathbf{\hat{r}}_{\mathrm{FM}}$ and the normal to the MC plane $\mathbf{\hat{n}}_{\mathrm{FM}}$ in the lower corona are determined then as
\begin{equation}\label{eq:}
	\mathbf{\hat{r}}_{\mathrm{FM}}=\cos\theta_{\mathrm{FM}}\cos\phi_{\mathrm{FM}}\mathbf{\hat{e}_x}+\cos\theta_{\mathrm{FM}}\sin\phi_{\mathrm{FM}}\mathbf{\hat{e}_y}+\sin\theta_{\mathrm{FM}}\mathbf{\hat{e}_z},
\end{equation}
\begin{equation}\label{eq:}
	\mathbf{\hat{n}}_{\mathrm{FM}}=rotate([\mathbf{\hat{r}}_{\mathrm{FM}}\times\mathbf{\hat{e}_y}]\times\mathbf{\hat{r}}_{\mathrm{FM}},\mathbf{\hat{r}}_{\mathrm{FM}},\gamma_{\mathrm{FM}}),
\end{equation}
where we defined vector rotation operator $rotate$ as
\begin{equation}\label{eq:}
	rotate(\mathbf{\hat{v}},\mathbf{\hat{a}},\gamma)=\mathbf{\hat{v}}\cos\gamma+(\mathbf{\hat{v}}\cdot\mathbf{\hat{a}})(1-\cos\gamma)\mathbf{\hat{a}}+[\mathbf{\hat{a}}\times\mathbf{\hat{v}}]\sin\gamma,
\end{equation}
which rotates $\mathbf{\hat{v}}$ around $\mathbf{\hat{a}}$ by angle $\gamma$ counterclockwise.

After having determined the orientation of the CME close to the Sun we turn our attention to {\insitu} observations near 1~AU. The Grad-Shafranov reconstruction (GSR) \cite{Hu2002} is used for the estimation of the local direction of the invariant axis of MC and reconstruction of a slice of the MC. We use the modified version of GSR described in Isavnin {\etal}, (\citeyear{Isavnin2011}). The output we are interested in are the local direction of the invariant axis of the flux rope $(\theta_{\mathrm{GSR}},\phi_{\mathrm{GSR}})$
\begin{equation}\label{eq:}
	\mathbf{\hat{n}}_{\mathrm{FR}}=\cos\theta_{\mathrm{GSR}}\cos\phi_{\mathrm{GSR}}\mathbf{\hat{e}_x}+\cos\theta_{\mathrm{GSR}}\sin\phi_{\mathrm{GSR}}\mathbf{\hat{e}_y}+\sin\theta_{\mathrm{GSR}}\mathbf{\hat{e}_z},
\end{equation}
and the impact parameter $\rho_{\mathrm{GSR}}$. The impact parameter is the measure of the closest approach of the spacecraft to the invariant axis of the flux rope, calculated either as the distance of the closest approach (in astronomical units) or the distance of the closest approach divided by the radius of the flux rope cross-section. Knowing the position of the spacecraft which crossed the MC $(\theta_{\mathrm{SC}},\phi_{\mathrm{SC}})$
\begin{equation}\label{eq:}
	\mathbf{\hat{r}}_{\mathrm{SC}}=\cos\theta_{\mathrm{SC}}\cos\phi_{\mathrm{SC}}\mathbf{\hat{e}_x}+\cos\theta_{\mathrm{SC}}\sin\phi_{\mathrm{SC}}\mathbf{\hat{e}_y}+\sin\theta_{\mathrm{SC}}\mathbf{\hat{e}_z}
\end{equation}
and the impact parameter $\rho_{\mathrm{GSR}}$ we can estimate the vector pointing to the part of the of flux rope closest to the spacecraft trajectory using the following equation:
\begin{equation}\label{eq:dir_to_fr_1AU}
	\mathbf{\hat{r}}_{\mathrm{FR}}=rotate\left(\mathbf{\hat{r}}_{\mathrm{SC}},\mathbf{\hat{n}}_{\mathrm{FR}},2\arcsin\frac{sign(\mathbf{\hat{n}}_{\mathrm{FR}}\cdot\mathbf{\hat{e}_y})\rho_{\mathrm{GSR}}}{2}\right).
\end{equation}
The vector defined by Equation (\ref{eq:dir_to_fr_1AU}) lies in the MC plane. We can estimate the normal to the MC plane at 1~AU as
\begin{equation}\label{eq:}
	\mathbf{\hat{n}}_{\mathrm{GSR}}=\mathbf{\hat{n}}_{\mathrm{FR}}\times\mathbf{\hat{r}}_{\mathrm{FR}}.
\end{equation}

\begin{figure}
\centerline{
\includegraphics[width=0.5\textwidth,clip=]{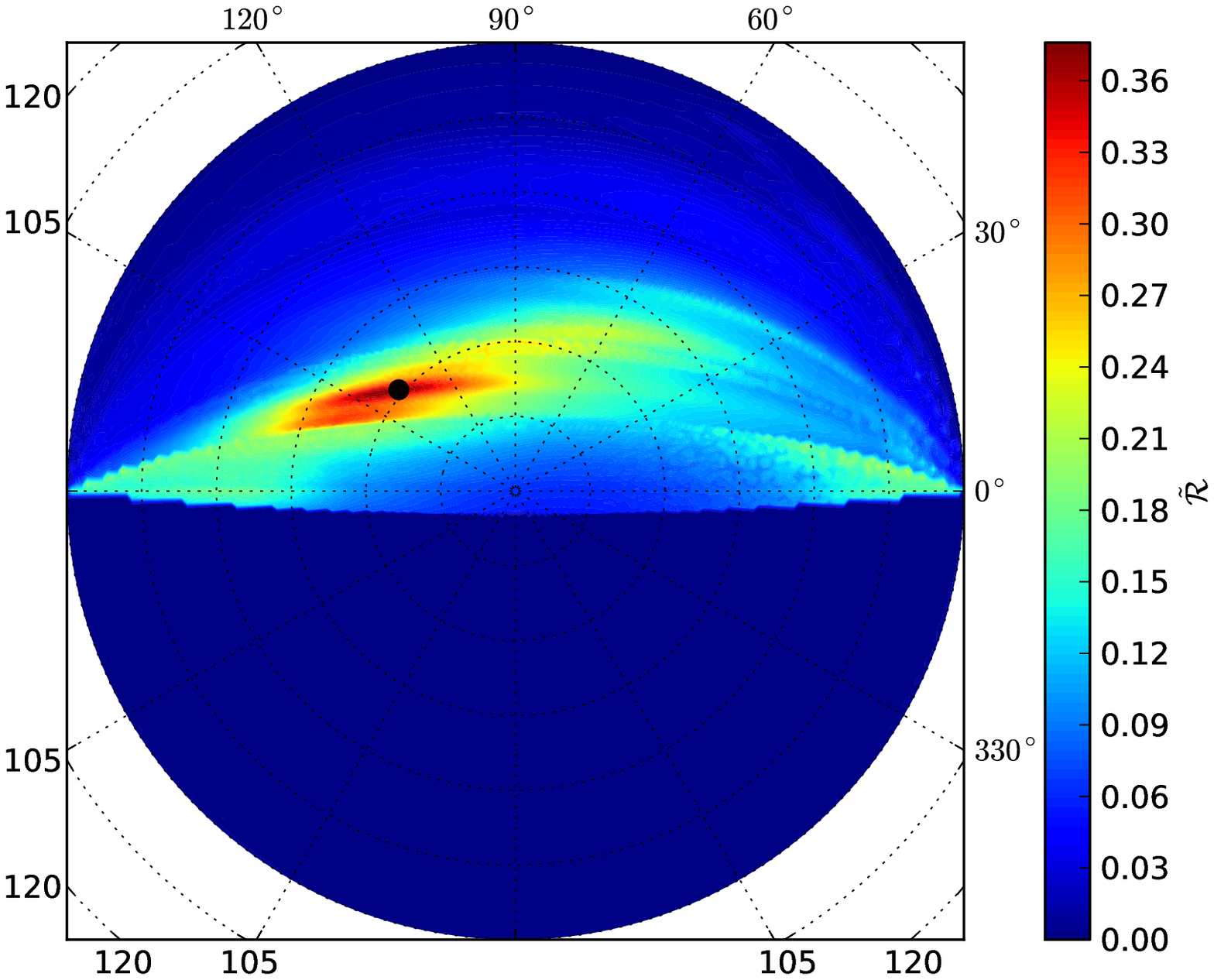}
\includegraphics[width=0.5\textwidth,clip=]{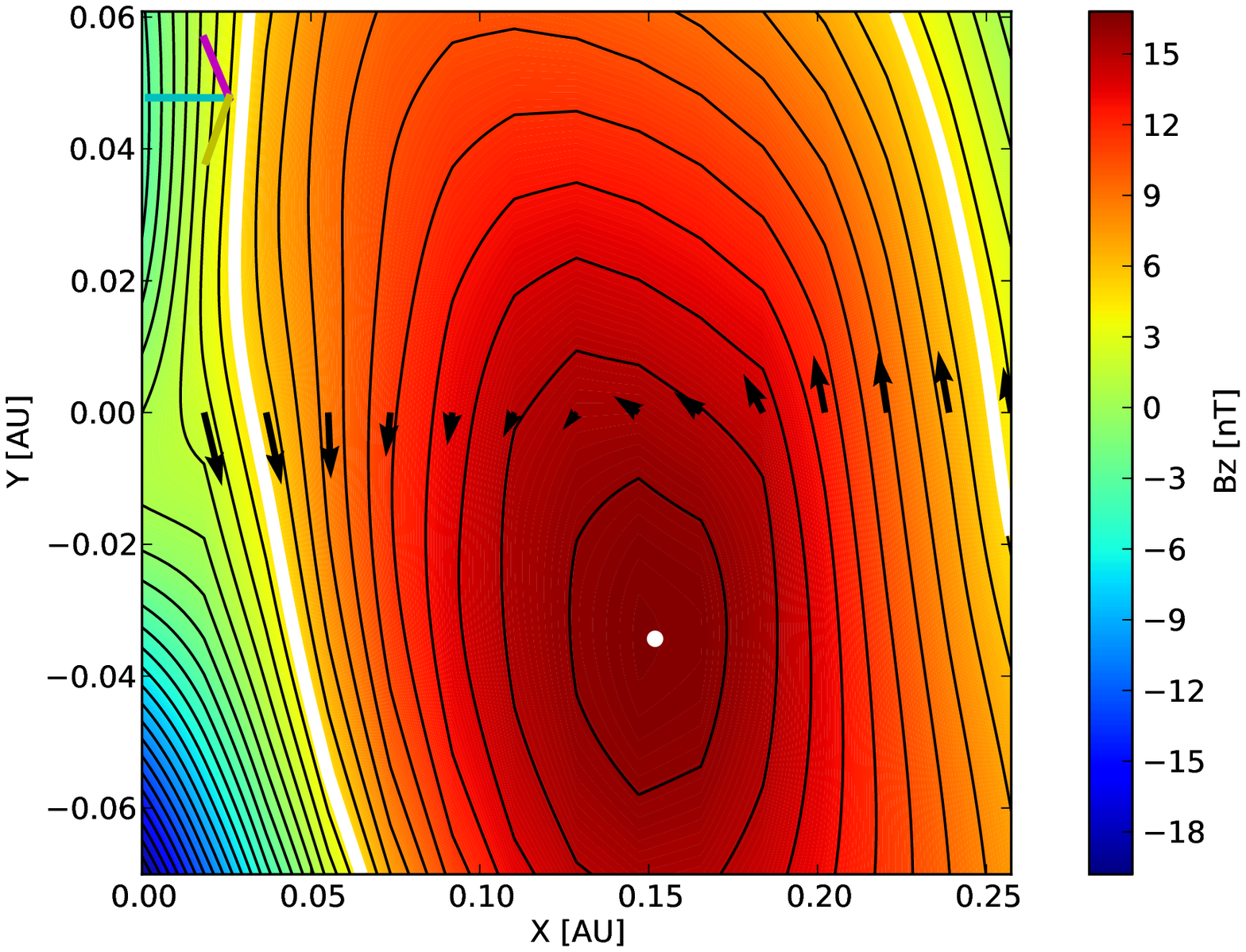}
}
\caption{Residual map (left) and magnetic field map (right) for the MC registered on 8 November 2010 by STEREO-B spacecraft. The projected RTN coordinate axes are $R_{RTN}$ (cyan), $T_{RTN}$ (magenta) and $N_{RTN}$ (yellow)}\label{fig:gsr_BzMap_example}
\end{figure}

The MC associated with the CME in example event was registered \textit{in situ} by the STEREO-B spacecraft on 8 November 2010. Figure~\ref{fig:gsr_BzMap_example} shows the residual map and the reconstructed magnetic field map for this event. The residual map shows the process of the search for the local direction of the invariant axis of the flux rope. It represents the hemisphere of all possible orientations of the axis and the direction with the minimal residue is the estimated invariant axis of the flux rope. The reconstructed magnetic field map is essentially the cross-section of the flux in the vicinity of the spacecraft trajectory. Black arrows show the magnetic field measured \textit{in situ} and projected onto the plane perpendicular to the invariant axis of the flux rope. Black contour lines denote the equipotential levels, where the absolute values of the vector potential are considered. The white dot represents the invariant axis of the flux rope. The thick white contour line shows the boundary of the unperturbed part of the flux rope.

\begin{figure}
\centerline{
\includegraphics[width=0.8\textwidth,clip=]{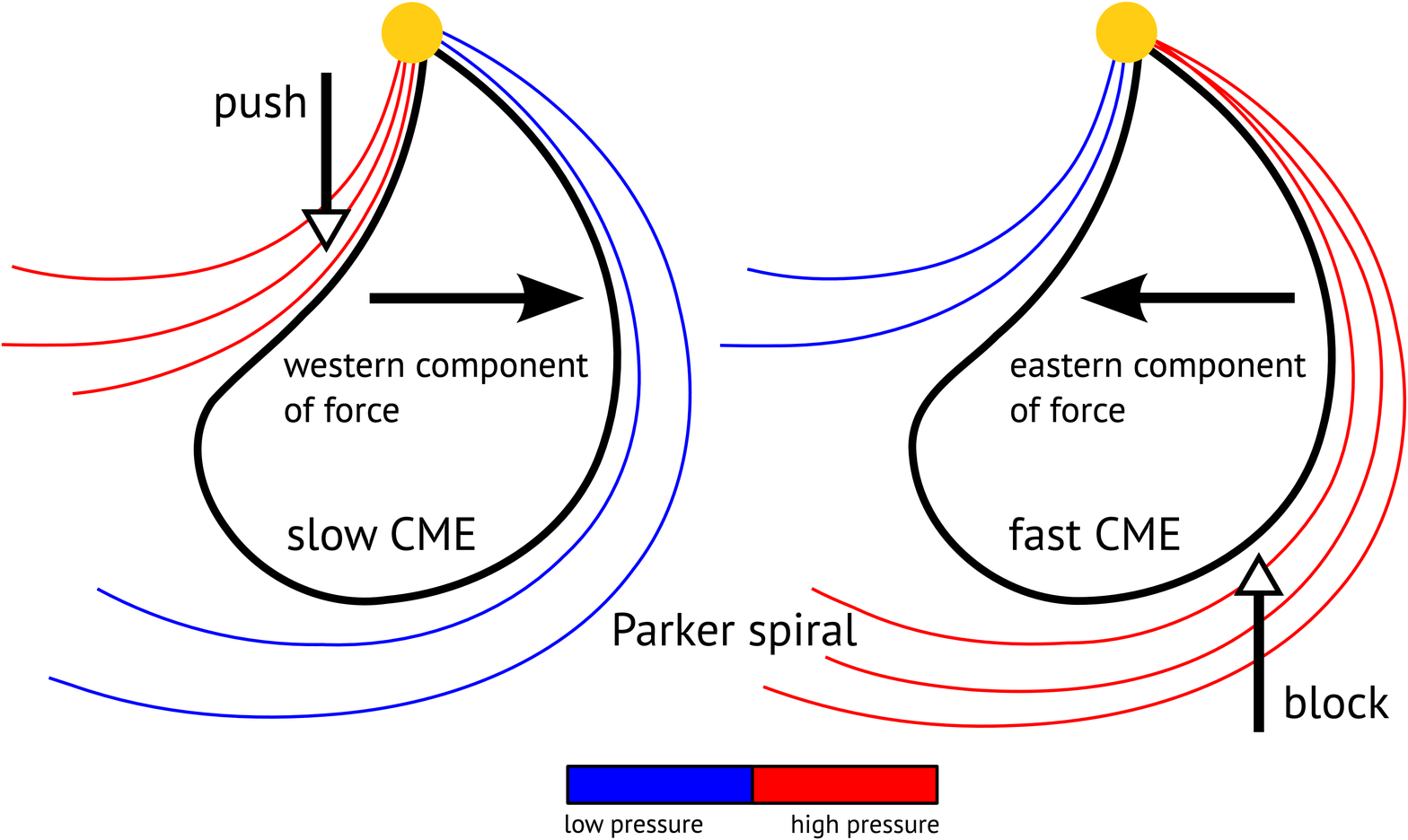}
}
\caption{Interaction of a slow (left) and fast (right) CME with Parker-spiral-structured solar wind (adapted from Wang {\etal}, 2004).}\label{fig:cme_parker_spiral}
\end{figure}

\inlinecite{Wang2004} introduced the kinetic interpretation of the longitudinal deflection of MCs which conforms with the statistics of observations of geoeffective MCs and their sources \cite{Wang2002}. According to these authors the longitudinal deflection of MCs is caused by the interaction of the MC with the Parker-spiral-structured solar wind (see Figure~\ref{fig:cme_parker_spiral}). If an MC propagates slower than the background solar wind it is pushed by the faster solar wind thus getting the westward component of the force. On the other hand, a fast MC is blocked by the slower background solar wind thus getting the eastward component of the force. Longitudinal deflection can be estimated using the following equation:
\begin{equation}\label{eq:long_defl}
	\Delta\phi=\Omega\left(\frac{1}{V_{\mathrm{MC}}}-\frac{1}{V_{\mathrm{SW}}}\right)\cdot\mathrm{1AU},
\end{equation}
where $V_{\mathrm{MC}}$ is the average speed of the MC propagation, $V_{\mathrm{SW}}$ is the average velocity of the background solar wind and $\Omega\approx2.7\times10^{-6}\mathrm{rad\,s^{-1}}$ is the angular velocity of the Sun's rotation. The positive value of $\Delta\phi$ represents westward deflection, while the negative value of $\Delta\phi$ shows eastward deflection of the MC. MCs which propagate with the velocity close to the background solar wind velocity do not experience longitudinal deflection on average. It should be noted that this is a rough estimate since the speed of the MC can change on its journey from the Sun to 1~AU. Since we are studying only events registered near the solar minimum, most of which are slow CMEs, the simplicity of Equation (\ref{eq:long_defl}) should not affect the results of the analysis significantly. For our example event the MC propagation velocity at 1~AU is slightly lower than the background solar wind. So we estimate the westward deflection using Equation (\ref{eq:long_defl}) to be $\Delta\phi=1^{\circ}$.

Longitudinal deflection is taken into account by rotating the initial MC direction $\mathbf{\hat{r}}_{\mathrm{FM}}$ by $\Delta\phi$ around $\mathbf{\hat{e}}_z$:
\begin{equation}\label{eq:}
	\mathbf{\hat{r}'}_{\mathrm{FM}}=rotate(\mathbf{\hat{r}}_{\mathrm{FR}},\mathbf{\hat{e}_z},\Delta\phi).
\end{equation}
Since $\mathbf{\hat{r}'}_{\mathrm{FM}}$ is directed along the global axis of the MC through its apex, the possible rotation of the MC around its axis will not affect $\mathbf{\hat{r}'}_{\mathrm{FM}}$. Then only latitudinal deflection can change $\mathbf{\hat{r}'}_{\mathrm{FM}}$:
\begin{equation}\label{eq:}
	\mathbf{\hat{r}''}_{\mathrm{FM}}=rotate(\mathbf{\hat{r}'}_{\mathrm{FM}},\mathbf{\hat{e}'_y},\Delta\theta),
\end{equation}
where $\mathbf{\hat{e}'_y}=\mathbf{\hat{e}_z}\times\mathbf{\hat{r}'}_{\mathrm{FM}}$ and $\Delta\theta$ is latitudinal deflection angle. The vectors $\mathbf{\hat{r}''}_{\mathrm{FM}}$ and $\mathbf{\hat{r}}_{\mathrm{FR}}$ both lie in the MC plane (see Figure~\ref{fig:mc_scheme}) and its normal at 1~AU can be estimated as
\begin{equation}\label{eq:normal1au}
	\mathbf{\hat{n}'}_{\mathrm{FM}}=\mathbf{\hat{r}}_{\mathrm{FR}}\times\mathbf{\hat{r}''}_{\mathrm{FM}}.
\end{equation}
Latitudinal deflection is determined as such an angle $\Delta\theta$ for which $\mathbf{\hat{n}}_{\mathrm{GSR}}\cdot\mathbf{\hat{n}'}_{\mathrm{FM}}=1$. The positive value of $\Delta\theta$ represents North-to-South deflection and negative value shows South-to-North rotation. For our example event we found the latitudinal deflection to be $\Delta\theta=-17^{\circ}$. The MC deflected from the Southern to the Northern hemisphere, crossing the helioequatorial plane.

After estimating $\Delta\phi$ and $\Delta\theta$ deflections the rotation can be calculated using the following equation:
\begin{equation}\label{eq:rotation}
	\Delta\gamma=arctan\left(\frac{\mathbf{\hat{r}}_{\mathrm{FR}}\cdot\mathbf{\hat{e}'_z}}{\mathbf{\hat{r}}_{\mathrm{FR}}\cdot\mathbf{\hat{e}'_y}}\right)-\gamma_{\mathrm{FM}},
\end{equation}
where $\mathbf{\hat{e}'_z}=\mathbf{\hat{r}''}_{\mathrm{FM}}\times\mathbf{\hat{e}'_y}$. The positive value of $\Delta\gamma$ represents counterclockwise rotation, while the negative value of $\Delta\gamma$ shows clockwise rotation of the MC around its global axis. Using Equation (\ref{eq:rotation}) we estimate a rotation by $\Delta\gamma=30^{\circ}$ for our example event. It should be noted that the FM fit contains no information about polarity or chirality of the flux rope. Thus, the estimated initial rotation angle of the flux rope has an ambiguity of 180 degrees. This implies that Equation (\ref{eq:rotation}) shows the change of the tilt angle of the MC plane, but not the real direction or full amount of rotation.

\section{Results}\label{s:results}

For the analysis we selected the MCs which were observed during the minimum following the Solar Cycle~23 and the rise of the Solar Cycle~24, \textit{i.e.} years 2008--2010. Such a choice is based on several considerations, \textit{i.e.} sufficiently large angular distance (at least 30--40 degrees) between the STEREO spacecraft which was launched in 2006 is required for the FM technique results to be reliable \cite{Thernisien2009}. The separation between the STEREO and SOHO spacecraft grew from $26^{\circ}$ to $87^{\circ}$ during that period. The minimum of solar activity is also characterised by a less dynamic structure of the heliospheric current sheet (HCS) and coronal holes which facilitate studies of interaction of an MC with these structures.

Similar to other flux rope fitting techniques, GSR works best for small impact parameter events (Isavnin {\etal}, \citeyear{Isavnin2011}), so we conducted analysis only for MCs which were crossed close to their invariant axis by the spacecraft at 1~AU. We have selected 15 events for our analysis, the results of analysis are summarized in Table \ref{tbl:results}.

\begin{figure}
\centerline{
\includegraphics[width=0.55\textwidth,clip=]{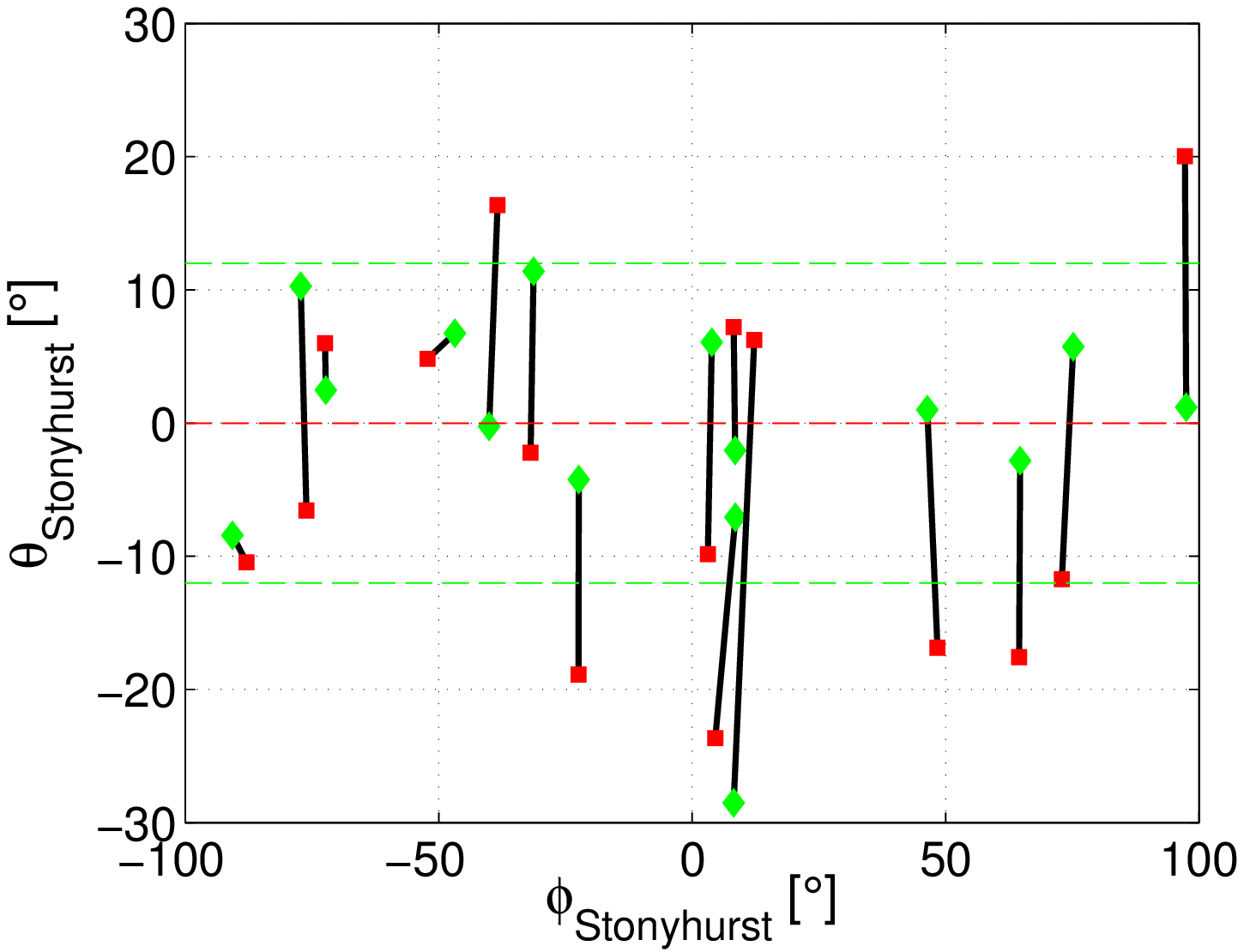}
\includegraphics[width=0.45\textwidth,clip=]{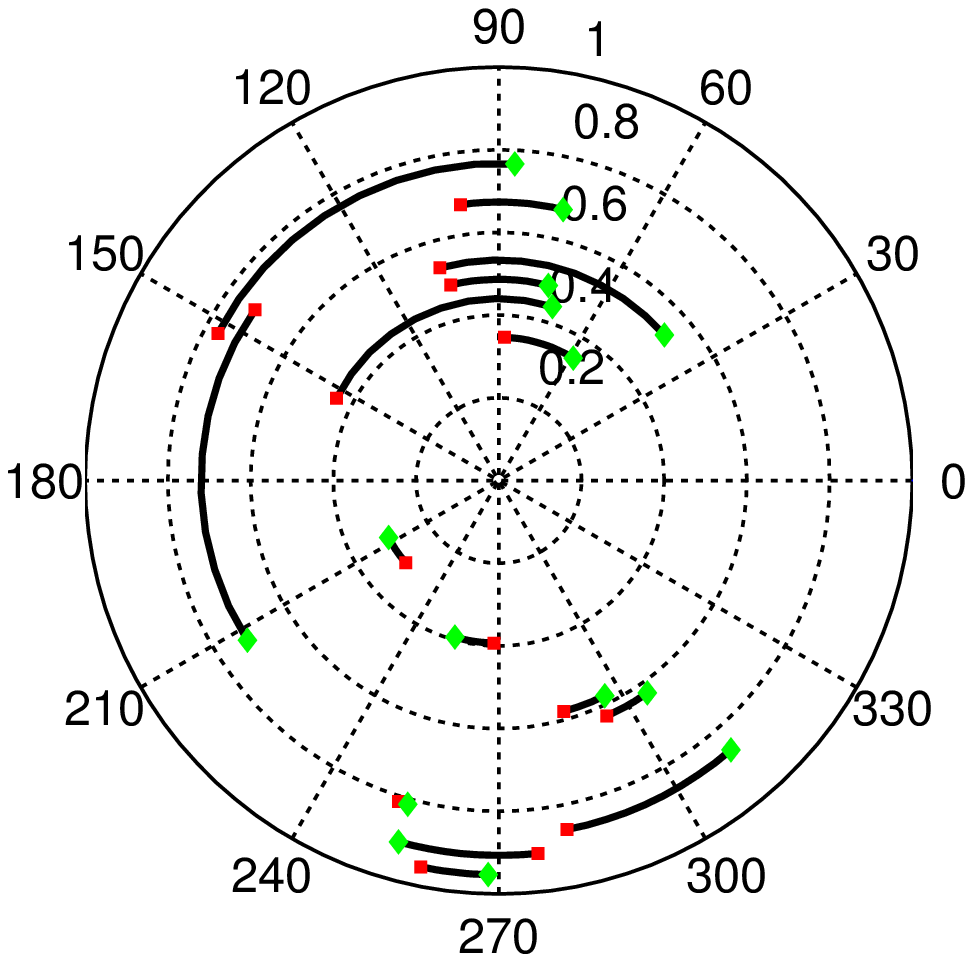}
}
\caption{Diagrams of deflection (left) and rotation (right) of the studied MCs. Red squares denote the direction of MC propagation (left) and the direction of the normal to the MC plane (right) close to the Sun, \textit{i.e.} in the initial position. Green diamonds denote the direction of MC propagation (left) and the direction of the normal to the MC plane (right) at 1~AU, \textit{i.e.} in the final position.}\label{fig:defl_rot_diagrams}
\end{figure}

The diagrams in Figure~\ref{fig:defl_rot_diagrams} visualize the estimated deflection and rotation experienced by the analyzed MCs when they propagated from the Sun to 1~AU. Also, the last three columns of Table \ref{tbl:results} give the longitudinal and latitudinal deflection angles (calculated from equations (\ref{eq:long_defl})--(\ref{eq:normal1au})) and the rotation angle estimated from Equation (\ref{eq:rotation}) for each event. In the left diagram of Figure~\ref{fig:defl_rot_diagrams} the $x$-components of the lines connecting the red squares and green diamonds show the amount of longitudinal deflection, while the $y$-components of the lines represent the amount of latitudinal deflection. In the right panel, the longer the angular distance covered by a curve between the red squares and green diamonds the more the MC has rotated during its interplanetary propagation. 

As shown by Figure~\ref{fig:defl_rot_diagrams} and Table \ref{tbl:results} for all studied events the longitudinal deflection was small, less than 6 degrees. The latitudinal deflection was clearly larger than the longitudinal, exceeding $10^{\circ}$ for the majority of events with the largest deflection of $35^{\circ}$ (event \#10). According to our study 14 out of 15 analyzed MCs deflected towards the helioequatorial plane, often crossing it. The only exception is event \#6 for which the deflection by $2^{\circ}$ away from helioequatorial plane was found. This event is also the one with the smallest amount of latitudinal deflection among the 15 events we studied. 

The amount of  rotation of the studied MCs ranged from events that experienced practically no rotation (event \#12) to events that rotated as much as 80 degrees (event \#4). The average absolute value of rotation angle for our data set was 31 degrees.

From the Figure~\ref{fig:defl_rot_diagrams}(left) it can be seen that the directions of propagation of 14 out of 15 MCs deflected to or stayed in the latitude range of $\theta_{Stonyhurst}=[-12^{\circ},12^{\circ}]$ after their travel from the Sun to 1~AU. Thus, MCs seem to align the direction of their expansion with the solar equatorial plane.

\section{Discussion}\label{s:discussion}

We have presented the first detailed study of the three-dimensional evolution of the orientation of MCs from the Sun to 1~AU. The analysis makes use of coronagraph images of CMEs and in-situ observations of erupted flux ropes and is capable of more precise identification of the three-dimensional orientation of individual flux ropes and their evolution than was possible in the past. This method utilizes the FM and GSR techniques for estimation of the flux rope orientation close to the Sun and at 1~AU, respectively. The approach is constrained by certain limitations on the quality of the studied events. The flux rope structure within the CME associated with the analyzed MC has to be clear enough to be easily distinguishable in the coronagraph images. The separation between the STEREO spacecraft has to be large enough (at least 30--40 degrees) to get three-dimensional representation of the CME and hence a better fit of FM. The impact parameter of the MC observed at 1~AU has to be small to estimate the orientation of MC at 1~AU more precisely using GSR technique, which works better for small impact parameters.

Based on these considerations we have selected 15 clear events during the years 2008--2010 and used our method to study their three-dimensional evolution from the Sun to 1~AU. The events were observed during the minimum following Solar Cycle~23 and the rise of Solar Cycle~24. Our analysis shows that MCs tend to deflect towards the solar equatorial plane on their journey from the Sun to 1~AU. This result is in agreement with previous statistical studies by MacQueen {\etal} (\citeyear{MacQueen1986}), \inlinecite{Plunkett2001} and Cremades {\etal} (\citeyear{Cremades2006}), though in our work we were able to calculate the evolution of MCs' orientation directly from the multi-spacecraft observations. In this study we have not considered the source regions of CMEs but we started tracking MCs from 2~$R_{\odot}$, hence we showed that the latitudinal deflection of MC can happen not only in the lower corona, but the orientation of MC continues to evolve all the way to 1~AU. A possible reason for the latitudinal deflection may be the kinematic interaction between CMEs and fast solar wind.

The studied events showed very little longitudinal deflection. As discussed in the Introduction, \inlinecite{Wang2002} observed an obvious East-West asymmetry in the source region distribution of geoeffective halo CMEs. While the sample of \inlinecite{Wang2002} covers the rising phase of solar activity and solar maximum, our study period coincides with relatively low solar activity conditions, and thus most of our CMEs were slow and embedded into solar wind of the speed close to the CME speed, \textit{i.e.} were not pushed by fast solar wind or blocked by slow solar wind. Thus, it is possible that our data set represents events that were not much influenced by the ambient solar wind flow. We also expect that forward modeling based on the coronagraph data gives a more reliable estimate of the CME propagation direction than the analysis of their source regions based on the solar disk observation. Our results are also consistent with the analysis by \inlinecite{Rodriguez2011} who found that the predictions of ICME detections based on the forward modeling of STEREO/COR2 data matched well with the actual {\insitu} observations. Our results thus imply that at least near solar minimum the CME propagation direction in longitude can be predicted accurately based on the coronagraph data and this direction does not change significantly from a few tens of solar radii from the Sun to 1~AU. It should be noted that although Equation (\ref{eq:long_defl}) is a rough estimate of the longitudinal deflection, it does not affect the accuracy of the presented technique dramatically. For instance, introducing an error $\delta\Delta\phi=10^{\circ}$ to the longitudinal deflection estimate leads to the average error $\delta\Delta\theta=3^{\circ}$ of latitudinal deflection estimate and $\delta\Delta\gamma=0.1^{\circ}$ of rotation angle estimate for the studied MCs.

The rotation of the studied MCs could be caused by the disconnection of one of the flux rope footpoints early in the eruption (Vourlidas {\etal}., \citeyear{Vourlidas2011}, Nieves-Chinchilla {\etal}., \citeyear{NievesChinchilla2012}), interaction with large-scale magnetic structures in the solar wind and HCS \cite{Yurchyshyn2008}. Analysis of these possibilities in relation to each event will be the subject of our upcoming research.

\begin{landscape}
\begin{table}[!ht]
\begin{center}
\begin{tabular}{r || r | r | r | r || r | r | r | r | r | r | r || r | r | r}
 \multirow{2}{*}{\#}     & \multicolumn{4}{c||}{CME}    & \multicolumn{7}{c||}{MC}       & \multicolumn{3}{c}{$\Delta$ orientation} \\
   &       date & 
    $\phi_{\mathrm{FM}}$ &
	       $\theta_{\mathrm{FM}}$ &
	 		        $\gamma_{\mathrm{FM}}$ &       date &   SC & 
	 		        							   $\phi_{\mathrm{SC}}$ &
   						 		           			  	 $\theta_{\mathrm{SC}}$ &
   						 		           						   $\phi_{\mathrm{GSR}}$ &
     															       	 $\theta_{\mathrm{GSR}}$ & 
     															       	 	    $\rho_{\mathrm{GSR}}$ [AU] &
   						 		           												             $\Delta\phi$  &
 						 		           												 		 		     $\Delta\theta$ &
   						 		           												 		 			 	   	  $\Delta\gamma$ \\ 
 \hline
 1 & 2008-06-02 & -31.85 &  -2.23 & -48.49 & 2008-06-06 &  STB & -24.97 & -3.77 & -152.9 &  58.0 & -0.0045 & -0.58 & -13.62 & -14.20 \\
 2 & 2008-07-07 & -22.39 & -18.88 &  -2.20 & 2008-07-10 &  STB & -27.20 &  0.10 &   88.3 & -28.6 &  0.0195 & -0.05 & -14.65 & -29.16 \\
 3 & 2008-08-31 &   3.10 &  -9.85 &  -1.64 & 2008-09-03 & \textit{Wind} &   0.08 &  7.18 & -107.4 &  12.9 &  0.0085 & -0.77 & -15.92 & -14.00 \\
 4 & 2008-12-12 &   8.19 &   7.20 &  63.11 & 2008-12-17 & \textit{Wind} &   0.08 & -1.17 &  115.7 & -15.8 & -0.0279 & -0.31 &   9.26 & -80.26 \\
 5 & 2008-12-27 & -38.37 &  16.37 &  13.74 & 2008-12-31 &  STB & -45.20 &  3.40 &   24.5 & -12.7 &  0.0264 &  1.66 &  16.62 & -28.03 \\
 6 & 2009-09-27 & -52.16 &   4.82 &  15.49 & 2009-10-02 &  STB & -56.73 &  5.60 &   75.1 & -42.7 & -0.0031 & -5.38 &  -1.92 & -64.17 \\
 7 & 2010-01-15 & -72.40 &   6.00 &  15.69 & 2010-01-20 &  STB & -69.07 &  3.90 & -180.0 & -25.8 & -0.0045 & -0.15 &   3.54 &  10.54 \\
 8 & 2010-02-01 &  64.54 & -17.59 &  24.64 & 2010-02-05 &  STA &  64.10 & -6.20 &  -51.9 & -31.2 & -0.0163 & -0.16 & -14.76 &  10.31 \\
 9 & 2010-04-03 &   4.60 & -23.66 &   7.93 & 2010-04-05 & \textit{Wind} &  -0.24 & -6.30 &  140.6 &  -3.4 &  0.0144 & -3.93 & -16.59 & -21.31 \\
10 & 2010-05-23 &  12.25 &   6.23 &  54.98 & 2010-05-28 & \textit{Wind} &   0.03 & -1.02 &   67.6 & -60.1 &  0.0007 &  4.05 &  34.75 &  67.43 \\
11 & 2010-05-27 &  72.92 & -11.73 &  62.35 & 2010-05-31 &  STA &  71.70 &  6.50 &  132.7 &   0.1 &  0.0165 & -2.28 & -17.46 & -65.27 \\
12 & 2010-06-13 &  97.22 &  20.04 & -17.36 & 2010-06-16 &  STA &  73.68 &  7.30 &   -3.4 &  15.2 & -0.0078 & -0.25 &  18.85 &   1.66 \\
13 & 2010-11-04 & -76.07 &  -6.56 &  11.06 & 2010-11-08 &  STB & -82.20 &  6.40 & -139.0 & -35.3 &  0.0341 &  1.11 & -16.84 &  29.69 \\
14 & 2010-12-12 &  48.40 & -16.87 &   6.02 & 2010-12-15 &  STA &  85.20 & -7.30 &   10.1 &  10.9 & -0.0158 &  2.00 & -17.86 & -21.52 \\
15 & 2010-12-12 & -87.88 & -10.46 & -11.40 & 2010-12-17 &  STB & -87.30 &  7.30 & -166.0 &   0.2 &  0.0158 &  2.78 &  -2.02 &  -9.85 \\
\end{tabular}
\end{center}
\caption{The results of the analysis of MC deflection for 15 events. The CME part of the table represents the results of FM analysis at $5-20R_{\odot}$, MC part of the table represents the results of GSR analysis at 1~AU, the $\Delta$ orientation represents the estimated deflection angles. The columns from left to right are as follows: \# -- event number, date -- date the CME event start, $\phi_{\mathrm{FM}}$, $\theta_{\mathrm{FM}}$ -- direction of the CME propagation, $\gamma_{\mathrm{FM}}$ -- initial rotation angle of the CME, date -- date of MC registration at 1~AU, SC -- spacecraft which observed the MC at 1~AU, $\phi_{\mathrm{SC}}$, $\theta_{\mathrm{SC}}$ -- coordinates of the spacecraft, $\phi_{\mathrm{GSR}}$, $\theta_{\mathrm{GSR}}$ -- direction of the invariant axis of the MC, $\rho_{\mathrm{GSR}}$ -- impact parameter, $\Delta\phi$ -- longitudinal deflection (positive for East-to-West), $\Delta\theta$ -- latitudinal deflection (positive for North-to-South), $\Delta\gamma$ -- change of the rotation angle (positive for counterclockwise).}
\label{tbl:results}
\end{table}
\end{landscape}

%

%

%
\begin{acks}
The work of A. Isavnin and E. Kilpua was supported by the Academy of Finland. The work of A. Vourlidas is supported by NASA contract S-136361-Y to the Naval Research Laboratory. LASCO was constructed by a consortium of institutions: NRL (USA), MPI fur Aeronomie (Germany), LAS (France) and University of Birmingham (UK). The SECCHI data are produced by an international consortium of the NRL, LMSAL and NASA GSFC (USA), RAL and University of Birmingham (UK), MPS(Germany), CSL (Belgium), IOTA and IAS (France).
\end{acks}

%
%
\bibliographystyle{spr-mp-sola}
\bibliography{bibliography.bib}  
%
%
%
%

\end{article} 
\end{document}